\documentclass[12pt,a4paper,twoside]{article}        
\usepackage[]{babel}                  
\usepackage[cp1251]{inputenc}                        
\usepackage{mmgart}                                  
\begin{document}

\setcounter{page}{1}                                
\thispagestyle{empty}                                
\begin{heading}                                      
{Volume\;0,\, N{o}\;0,\, p.\,13 -- 24\, (2014)}      
{Special issue}                                      
\end{heading}                                        

\begin{Title}
Modeling of  $^6_\Lambda$He hypernucleus  within configuration space Faddeev approach
\end{Title}

\begin{center}
\Author{a}{I. Filikhin},
\Author{}{V. M. Suslov}
\and
\Author{}{B. Vlahovic}
\end{center}


\begin{flushleft}
\Address{}{Department of Physics, North Carolina Central University, 1801 Fayetteville Street, Durham, NC 27707, USA
}

\Email{$^a$\,ifilikhin@nccu.edu}
\end{flushleft}

\Headers{I. Filikhin et al.}{Cluster calculations
}


\Thanks{
This work is supported by the NSF (HRD-0833184) and NASA (NNX09AV07A).
}

\Thanks{\mbox{}\\
\copyright\,The author(s) 2013. \ Published by Tver State University, Tver, Russia}
\renewcommand{\thefootnote}{\arabic{footnote}}
\setcounter{footnote}{0}

\Abstract{The cluster
$^4\rm He+\Lambda+\rm n$ model is applied to describe
the $^6_\Lambda$He  hypernucleus. The consideration is
based on the configuration space Faddeev equations for a system of non-identical particles.
A set of the pair potentials includes 
the OBE simulating (NSC97f) model for the $\Lambda \rm n$ interaction and the phenomenological potentials for the $\alpha\Lambda$  and 
 $\alpha \rm n$ interactions.
We calculated  energies of spin (1$^-$,2$^-$)  doublet. 
For the 2$^-$ excitation energy, the obtained value is 0.18~MeV. The hyperon binding energy of the  bound 1$^-$ state is less than
the experimental value, which may be an evidence for
violation of the exact three-body cluster structure. 
}
\\
\noindent
\\


\newpage                               
\renewcommand{\baselinestretch}{1.1}   


\section{Introduction}

The spin doublet 1$^-$, 2$^-$  of the $^6_\Lambda$He hypernucleus is of a great interest for testing the theoretical models of the hyperon-nucleon interaction, in particular, for the study of spin-dependence of the interaction \cite{M, Mo, H1996}. Especially, this study is important in relation to the recently observed the $^7_\Lambda$He \cite{7HeL}, $^6_\Lambda$H hypernuclei \cite{6HL} and possible bound $\Lambda$n system \cite{BMP}.

The experimental value is  known for  hyperon binding energy of the $^6_\Lambda$He ground state \cite{Exp}. 
The binding energy
$E_n$ of the $^6_\Lambda$He  with respect to the two-particle decay accompanied by the separation of the
loosely bound neutron was measured to be $E_n$=0.17$\pm$0.10 MeV. The parameters of the  2$^-$ resonance state  are not measured  yet.

Theoretical considerations for the states have been done by Motoba et al. \cite{Mo} and Hiyama et al. \cite{H1996} using the three-body cluster $\alpha \Lambda $n model. Indirect prediction for the (1$^-$, 2$^-$) energy spacing was given in our  work \cite{FSV2005}. 
In Ref.  \cite{Mo}, it was found  that the spin-orbital component of the wave function can be represented as $0.998(10)^{1}_{1}$ for the 2$^-$ state and $0.688(10)^{0}_{1}+0.309(10)^{1}_{1}$ for the 1$^-$ state. Here, 
the $LS$-coupling scheme is presented by the following form $(l\lambda)^{S}_{L}$ with the definitions: $l$ is orbital momentum of the core nucleus ($^5$He), $\lambda$ is orbital momentum of the hyperon, and $S$ and $L$ are the total spin and total orbital momentum. Thus, the  state 2$^-$ is pure triplet spin state and state 1$^-$ is superposition of triplet and singlet spin states.
Note, however, that this calculation  underestimated 
the  binding energy of $^6_\Lambda$He for the potentials chosen.
In Ref.  \cite{H1996} a similar treatment was performed with 
 modified  potentials, which  overestimated the system. A new corrected value  for the   binding energy of  $^6_\Lambda$He was reported in Ref. \cite{HOKY14}. The value is within the experimental errors, but it also slightly overestimates the system.

In the present work,  the $^6_\Lambda$He hypernuclus is modeled as a cluster $\alpha \Lambda $n system. 
We use the potentials proposed for the $\alpha$n and $\Lambda$n  interactions in Refs. \cite{pot-an,GF} to evaluate
energies of the  (1$^-$, 2$^-$) spin doublet. The model 
does not include  the $\Lambda\rm N$--$\Sigma\rm N$
coupling  that may be strong, as  it was indicated in \cite{Mil06}.
Our calculations  are based on the configuration-space Faddeev equations for a system of three non-identical particles. 
 We  compare our numerical results with those of the \cite{H1996, HOKY14} calculations
  and with the experimental data. 
 
\section{Model}

The   cluster  $\alpha \Lambda n$ system may be described by the Faddeev equations \cite{fad} which are given as follows
\begin{equation} \label{EQ1}
\begin{array}{l} {(H_{0} +V_{\Lambda n} -E)U_1=V_{\Lambda n} (U_2+U_3),} \\ {(H_{0} +V_{\alpha n} -E)U_2=V_{\alpha n} (U_1+U_3),} \\
{(H_{0} +V_{\alpha \Lambda} -E)U_3=V_{\alpha \Lambda} (U_1+U_2).} 
\end{array}
\end{equation}
The total wave function of the system $\Psi$ is decomposed in the sum of the Faddeev components $U_1$, $U_2$, and
$U_3$ corresponding to the $(\Lambda n)\alpha $,  $(\alpha n)n$, and $(\alpha \Lambda)n$
types of
rearrangements, respectively:
\[\Psi =U_1+U_1+U_3.\]
$V_{\Lambda n}$, $V_{\alpha n}$, and $V_{\alpha \Lambda}$ are pair potentials of interactions between $\Lambda -$n,  $\alpha -$n, and $\alpha -\Lambda$.

The differential form of the Faddeev equations means that each Faddeev component is expressed in its own Jacobi coordinates  ${\bf x}_i,{\bf y}_i$, $i$=1,2,3 
\[ U_i=U_i({\bf x}_i,{\bf y}_i).\]
In Eq. (\ref{EQ1}), $H_0$ is differential operator $H_0=-\frac{\hbar^2}{m}(\Delta_{{\bf x}_i}+\Delta_{{\bf y}_i})$. 
 The Jacobi vectors are linearly related by the orthogonal transformation
$$
  \left(
  \begin{array}{c}
     \bf{x}_i \\ \bf{y}_i
  \end{array}
  \right)=
  \left(
  \begin{array}{rl}
      C_{ij} & S_{ij} \\
     -S_{ij} & C_{ij}
  \end{array}
  \right)
  \left(
  \begin{array}{c}
     \bf{x}_j \\ \bf{y}_j
  \end{array}
  \right) \ ,\ \ \ C^2_{ij} + S^2_{ij} = 1,
$$
where
$$
C_{ij}=-\sqrt{\frac{m_im_j}{(M-m_i)(M-m_j)}} \ \ , \ \ S_{ij}
= (-)^{j - i} \mbox{sign}(j - i)\sqrt{1-C^{2}_{ij}}.
$$
$M$ is the total mass of the system and $m_i$ is the mass of the $i$-th particle.

The $LS$-coupling scheme is used for partial wave analysis
of the equations (\ref{EQ1}). 
A quantum state in LS basis is defined by  a set of quantum numbers
$\{(l\lambda) L (\sigma s) S\}\equiv \alpha$.
In this basis the spin-angular eigenfunctions have the form
$$
W_{\alpha}({\bf \hat{x},\hat{y}})=<{\bf \hat{x},\hat{y}}|\alpha>=
\Big[ [Y_l({\bf \hat{x}})\otimes Y_{\lambda}({\bf \hat{y}})]^{LL_z}\otimes
[\sigma \otimes s]^{SS_z}\Big]^{J,J_z}.
$$
For convenience, one may use the bipolar harmonics:
$$
{\cal Y}^{LL_z}_{l\lambda}({\bf\hat{x},\hat{y}})=[Y_l({\bf \hat{x}})\otimes Y_{\lambda}({\bf \hat{y}})]^{LL_z}=\sum_{m_l,m_{\lambda}}
C^{LL_z}_{lm_l\lambda m_{\lambda}}
Y_{lm_l}({\bf \hat{x}})Y_{\lambda m_{\lambda}}({\bf \hat{y}}).
$$
The explicit form of the eigenfunction is 
$$
W_{\alpha}({\bf\hat{x},\hat{y}})=\sum_{L_z,S_z,\sigma_z,s_z}
C^{JJ_z}_{LL_zSS_z}C^{SS_z}_{\sigma\sigma_zss_z}
{\cal Y}^{LL_z}_{l\lambda}({\bf\hat{x},\hat{y}})
\chi^{\sigma}_{\sigma_z}\chi^{s}_{s_z}.
$$
Correspondingly the Faddeev component expanded in the basis
$|\alpha>$ is
$$
U_i({\bf x_i,y_i})=\sum_{\alpha} \frac{\Phi_{\alpha}(x_i,y_i)}{x_i,y_i}
W_{\alpha}({\bf \hat{x_i},\hat{y_i}}).
$$

The sum over $\alpha=\{(l\lambda) L (\sigma s) S\}$ is restricted by the coupling
triangular condition and the parity conservation condition
$(-1)^{\lambda+l}=(-1)^{\pi}$. 
The final set of the equations for the Faddeev components (\ref{EQ1}) is
\[
\Big[ -\frac{\hbar^2}{m}(\partial_{x^2_i}+\partial_{y^2_i})+
v_{\alpha}^{cen.}(x_i,y_i)-E) \Big]\Phi_{\alpha}(x_i,y_i)=
-\sum_{\beta}v_{\alpha\beta}(x_i)\Big[\Phi_{\beta}(x_i,y_i)\]
\begin{equation}
\label{EQ2}
+\frac12\int^1_{-1} du 
\sum_{\gamma}\Big(\frac{x_iy_i}{x_jy_j}h_{\beta\gamma}(x_j,y_j,u)
\Phi_{\gamma}(x_j,y_j)+
\frac{x_iy_i}{x_ky_k}h_{\beta\gamma}(x_k,y_k,u)
\Phi_{\gamma}(x_k,y_k)\Big)\Big],\ \  i\ne j\ne k,
\end{equation}
where the centrifugal potential is
$$
v^{cen.}_{\alpha}(x_i,y_i)=\frac{\hbar^2}{m}\Big[\frac{l(l+1)}{x_i^2}+
\frac{\lambda(\lambda+1)}{y_i^2}\Big].
$$
The $v_{\alpha\beta}(x_i)$ are matrix elements of nuclear potentials.
General form of the $h_{\alpha'\alpha}(x_j,y_j,u)$ functions has the form 
$$
\begin{array}{l}
\displaystyle
h_{\alpha'\alpha}(x_j,y_j,u)=
(-)^{l'+\lambda'+L}{\prod_{l\lambda}}^2\prod_{l^{\prime}\lambda^{\prime}}(2l)!(2\lambda)!
\sum_{\displaystyle{l_1+l_2=l_{_{_{_{\hspace{-1.8cm}\displaystyle \lambda_1+\lambda_2=\lambda}}}}}}
(-)^{\lambda_1}\frac{x_i^{l_1+\lambda_1}y_i^{l_2+\lambda_2}}{x^{l}_jy^{\lambda}_j}
\\
\displaystyle
\frac{C_{ji}^{l_1+\lambda_2}S_{ji}^{l_2+\lambda_1}}
{\sqrt{(2l_1)!(2l_2)!(2\lambda_1)!(2\lambda_2)!}}
\sum_{k}(-)^k(2k+1)P_{k}(u)\sum_{l^{\prime\prime}\lambda^{\prime\prime}}
C^{l^{\prime\prime}0}_{l_10 \lambda_10} C^{\lambda^{\prime\prime}0}_{l_20 \lambda_20}
C^{l^{\prime\prime}0}_{l'0 k0}C^{\lambda^{\prime\prime}0}_{\lambda'0 k0}
\\
\displaystyle
\left \{ \begin{array}{ccc}
    l'       & l^{\prime\prime}       & k\\
    \lambda^{\prime\prime} & \lambda' & L\\
    \end{array} \right \}    
\left \{ \begin{array}{ccc}
    l_1  & \lambda_1 & l^{\prime\prime} \\
    l_2  & \lambda_2 & \lambda^{\prime\prime}\\
    l    & \lambda   & L
    \end{array} \right \},
\end{array}
$$
where $\prod_{l\lambda}=\sqrt{(2l+1)(2\lambda+1)}$ and $0\le k\le (\lambda^{\prime\prime}+l^{\prime\prime}+\lambda+l)/2$.

In the model, the  potential $\alpha$n
 is the only one  that is written as  a sum of the central and the spin-orbit parts. The matrix elements of the spin-orbit potential $V^{so}_{\alpha n}$ are given as 
$$
 V^{so}_{\alpha n}(x)=
\frac{2L+1}2\sum_{j=l \pm 1/2}(2j+1)
\left \{ \begin{array}{ccc}
    J & L & 1/2  \\
    l & j & \lambda
    \end{array} \right \}^2
$$
\[
\times (j(j+1)-l(l+1)-3/4)v_{so}(x),\]
where 
$v_{so}(x)$ is a coordinate part of the $\alpha$n spin-orbit potential.
In the $LS$ basis  the  $\alpha$n potential
is represented by diagonal matrix with diagonal elements:
$v^l_{\alpha n}(x)=v^l_{c}(x)+V^{so}_{\alpha n}(x)$, where $v^l_{c}(x)$ is central $l$-wave partial component of the $\alpha$n potential.

We consider spin-orbital configurations in the system corresponding to 
the total angular momentum $J^\pi$ equal to 1$^-$ and 2$^-$.
The $LS$ basis
allows to restrict 2$^-$ model space  to the states with the total spin $S=1$ (when the spin projections of hyperon and nucleon are parallel). 
For 1$^-$ state, 
the possible spin-orbital momentum configurations 
are represented with a set of $S=0$ and $S=1$. 

For both considered states 1$^-$ and 2$^-$ , the total orbital momentum $L^\pi$=1$^-$ has the following combinations of the subsystem orbital momenta
\begin{eqnarray}
\{({\ell}_{\alpha \Lambda},{\lambda}_{(\alpha\Lambda)-\rm n})\}& = & (0,1),(1,0),(1,2), \nonumber \\
\{({\ell}_{\alpha \rm n}, {\lambda}_{(\alpha \rm n)-\Lambda})\}& = & (0,1),(1,0),(1,2),(2,1), \nonumber \\
\{({\ell}_{\Lambda \rm n}, {\lambda}_{(\Lambda \rm n)-\alpha})\}& = & (0,1). \nonumber
\end{eqnarray}
Within our model, the $\alpha$n  interaction is constructed to reproduce the results of $R$-matrix analysis for $\alpha-$n scattering data \cite{N2007}. This potential simulates the Pauli exclusion principle for $\alpha n$ in the s-state with repulsive core. 
This new  $\alpha n$ potential was proposed in Ref. \cite{pot-an}. The parameters of the potential were obtained by modification of the 
potential given in Ref. \cite{CFJ89}.
Using this potential,  we  modeled the $^6$He nucleus as the $\alpha$nn cluster system. The results of the calculations reproduce well 
the experimental data for the low-lying spectrum of the nucleus \cite{FSV14}.
The potential is represented as one and two range Gaussian function:
\begin{equation}
\label{pot} V^l_{}(x) =
V^l_{rep}\exp(-\beta^l_{rep}x)^2-V^l_{att}\exp(-\beta^l_{att}x)^2\;\;.
\end{equation}
The parameters of the potential are presented in Tab. \ref{tab:1}.

\begin{table} [pt]
\caption{ Parameters of the $\alpha$n and
$\alpha\Lambda$ potentials (\ref{pot}). The pair angular momentum is
$l$.  $V_{\rm att}^l$ ($V_{\rm rep}^l$) and $V_{c}$ are given in MeV,
 $\beta_{\rm att}^l$ ($\beta_{\rm rep}^l$) in fm$^{-1}$, and
 $\alpha_{c}$ in
 fm$^{-2}$.
}
\label{tab:1}
\begin{tabular}{clccccc}
\hline
Potential & Component & $l$ &
$V_{\rm rep}^l$  & $\beta_{\rm rep}^l$ & $V_{\rm att}^l$ & $\beta_{\rm att}^l$ \\
\hline
$\alpha$n &  central   &0 & 50.0 & 1/2.3 & -- & -- \\
               &       &1 & 40.0 & 1/1.67& 63.0  & 1/2.3 \\
               &       &2 & --    & --   & 21.93 & 1/2.03 \\
     & spin-orbit & -- & -- & -- & 38.0  & 1/1.67   \\
\hline
$\alpha\Lambda$&  central   & 0,1,2,$\dots$  & -- & -- & 54.36 & 0.538  \\
\hline
\end{tabular}
\end{table}

An $\alpha\Lambda$ potential was proposed in Ref. \cite{8} to describe 
the $^9_\Lambda$Be hypernucleus within the cluster model $\alpha+\alpha+\Lambda$. 
The parameters of the potential are given in Tab. \ref{tab:1}.
The  chosen parameters allowed us to reproduce the energy of the ground state and the first excited state of $^9_\Lambda$Be,  simultaneously \cite{9}. 

We use the $\Lambda \rm N$ potential simulating the NSC97  model of barion-barion interaction \cite{RSY99}. This $s$-wave
potential is the three-range Gaussian \cite{HKM97} corrected in
\cite{GF} and has the following form:
\begin{equation}
V^{(2S+1)}(x)=\sum_i^3v_i^{(2S+1)}\exp(-\frac{x^2}{\beta_i^2})\;\;.
\label{eq:OBE}
\end{equation}
Here $S$ is spin of the pair ($S$=0 or 1). 
The values of the range parameters $\beta_i$ and of the singlet-
and triplet-strength parameters $v_i^{(2S+1)}$ are listed in table
\ref{tab:2}. In this table the parameter $\gamma$ was chosen so
that the potential (\ref{eq:OBE}) might reproduce the scattering
length and the effective range for the given model \cite{RSY99}.
In particular, singlet (triplet) scattering length is -2.5~fm (-1.75~fm).

\begin{table}
\caption{\label{tab:2}Parameters of the
 $\Lambda $n  potential (\ref{eq:OBE}).
The NSC97f model of barion-barion interaction is simulated by
choosing $\gamma^{(1)}$=1.0581 and $\gamma^{(3)}$=1.0499 \cite{GF}.}
\begin{tabular}{@{}cccc} \hline
$i$~ &~$\beta_i$ (fm)~&~$v_i^{(1)}$ (MeV)~&~$v_i^{(3)}$ (MeV)\\
\hline
 1~&~1.342~&~$-21.49$~&~$-21.39$~ \\
 2~&~0.777~&~$-379.1 \times \gamma^{(1)}$~&~$-379.1 \times \gamma^{(3)}$~ \\
 3~&~0.350~&~9324~&~11359~\\  \hline
\end{tabular}
\end{table}

\section{Results of calculations}

The  states 1$^-$ and 2$^-$  
of the cluster $\alpha \Lambda$n system 
are close to the two body $_{\Lambda }^{5} $He+n  threshold.
The lower level of the  doublet (1$^-$, 2$^-$) is a bound  
state. The upper member is a resonance state.
To evaluate energies of these states,
we solved numerically the bound state problem 
formulated by Eq. (\ref{EQ2}), applying the finite difference approximation
for radial variable and the spline collocation method for angular one.
The resulting matrix eigenvalue problem has been solved with the inverse iteration method. For the resonance state,
the method of analytical continuation in a coupling constant is used to calculate parameters of the resonance \cite{49}.
 The coupling constant is the depth of an additional non-physical three-body  potential \cite{KK05,50}. The potential has the form:
 $V_{3} (\rho )=-\delta \exp (-\alpha \rho ^{2} )$ with
 parameters $\alpha $, $\delta \ge $0 that can be varied. 
Here $\rho $ is hyper-radius of the tree-body system: $\rho ^{2} =x^{2}
+y^{2} $, where $x$, $y$ are the mass scaled Jacobi coordinates.
 For each resonance there exists a
 region \textbar $\delta $\textbar  $>$ \textbar $\delta _{0} $\textbar  where a resonance becomes a bound state.
 In this region we calculated 2$N$ bound state energies corresponding to 2$N$ values of $\delta $. The continuation
 of this energy set as a function of $\delta $ onto complex plane is carried out by means of the P\'ade approximant:
 $-\sqrt{E} =\sum _{i=1}^{N}p_{i}  \zeta ^{i} /(1+\sum _{i=1}^{n}q_{i} \zeta ^{i}  )$, where $\zeta =\sqrt{\delta -\delta_{0} } $.
 The complex value of the P\'ade approximant for $\delta $=0 gives the energy and width of the resonance:
 $E(\delta =0)=E_{r} +i\frac{\Gamma }{2} $.
 
\begin{figure}[!hb]
 \centering
   \includegraphics[scale=.2]{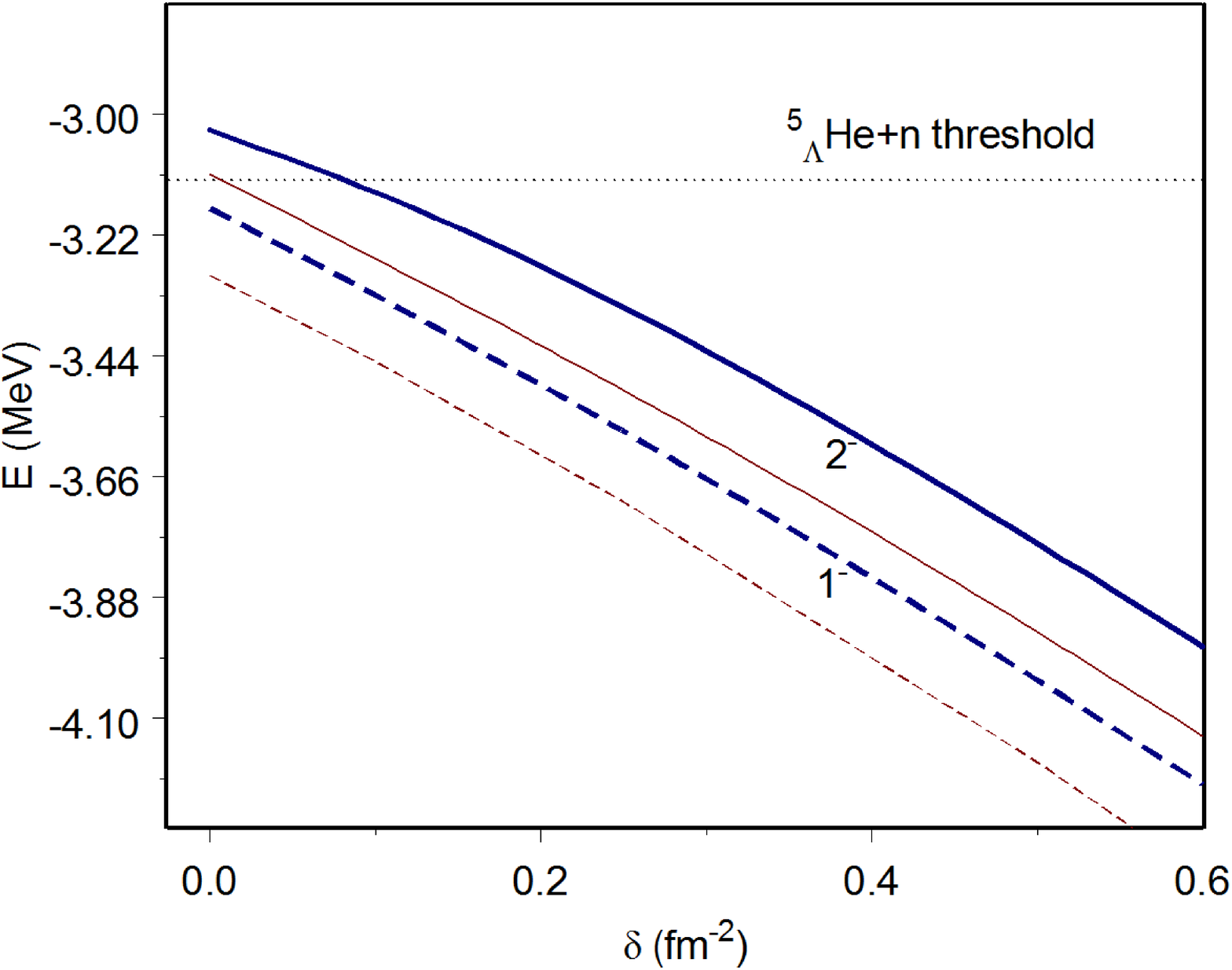}%
\caption {\small 
Real part of the P\'ade approximant for  $2^{-}$ state (solid line) of the $\alpha$+$\Lambda$+n  system.
Calculated resonance energy correspons to the P\'ade approximant value for $\delta$=0.  Energy is
measured from the $\alpha +n+n$ threshold. The  calculations for $1^{-}$ state are presented by  dashed line. 
 The range parameter $\alpha $ for used the three-body potential is 0.3~fm$^{-2}$.
The results that correspond to the calculations with the three-body force $V_{3bf}(\rho )=V_0 \exp (-\alpha_0 \rho ^{2} )$  are shown by the fine lines. The potential has parameters $V_{0}$=-0.05~fm$^{-2}$ and  $\alpha_0 $=0.2~fm$^{-2}$.
}\label{fig_1}
\end{figure}
  
 Results of the calculation are shown in Fig. \ref{fig_1}.
The calculated value for hyperon binding energy of $_{\Lambda }^{6} $He(1$^-$) is 0.65 MeV. The  resonance  energy of  $_{\Lambda }^{6} $He(2$^-$) is about 0.11 MeV. 
Since the experimental value for the $_{\Lambda }^{6} $He 
ground state was reported to be $E_\Lambda$=-0.17$\pm$~0.10~MeV, our cluster model underestimates the hypernucleus.
It is not surprising, taking into account the fact that  the similar
$\alpha$nn cluster system is also weakly bound (or is not bound) for the pair potentials under consideration. The cluster structure of such systems may be distorted at small distances between particles. Use of the three-body force  is proposed as an acceptable solution of the problem. 
Parameters of the three-body potential can be chosen
to reproduce experimental data (for instance, see our work \cite{FSV14}). The three-body force makes an equidistant  energy shift of spectral levels \cite{FSV14} (see also Fig. \ref{fig_1}). There could be  a total orbital momentum dependence \cite{FSV14} of the three-body potential. However, for the case
of the  doublet states (1$^-$, 2$^-$) of  the $\alpha\Lambda$n system, this dependence may be ignored due to the same orbital momentum for both states.
The three-body potential defined  as $V_{3bf}(\rho )=V_0 \exp (-\alpha_0 \rho ^{2} )$  with the parameters $V_{0}$=-0.05~fm$^{-2}$ and  $\alpha_0 $=0.2~fm$^{-2}$ allowed us to reproduce the experimental value for
the hyperon ground state energy, to be about  0.17$\pm$~MeV. The energy of the 2$^-$ resonance becomes close to  $^5_\Lambda$He+n threshold. This situation is shown in Fig. \ref{fig_1}. 
\begin{figure}[!hb]
 \centering
   \includegraphics[scale=.725]{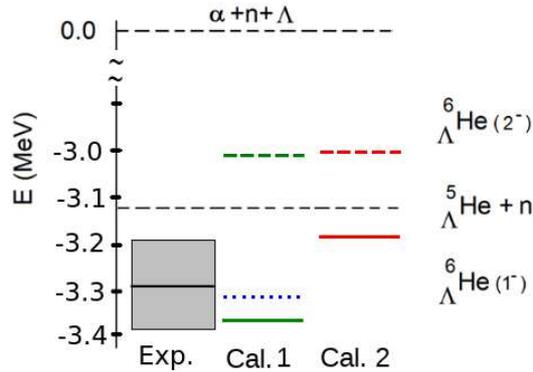}%
\caption {\small 
Spin-doublet (1$^-$,2$^-$) of the $^6_\Lambda$He nucleus.
Cal. 1 corresponds to the Refs. \cite{H1996} and \cite{HOKY14}. Cal. 2 are the present calculations. Exp. represents the experimental data 
with the error bar.
Solid (dashed) lines are   results calculated for the  ${1}^{-}$ 
bound state (${2}^{-}$ resonance). 
Doted line is the result for ${1}^{-}$ state from Ref. \cite{HOKY14}.
}\label{fig_2}
\end{figure}

Our prediction  for the (1$^-$,2$^-$) energies  can be directly compared with either  experimental data  or  other calculations.
In Fig. \ref{fig_2} we present our result together with the results of calculations from 
\cite{H1996} and \cite{HOKY14}.
In the model \cite{HOKY14},
a potential simulating  Nijmegen model NSC97f was used for the $\Lambda$n interaction. 
These calculations  resulted in the hyperon separation energy $B_\Lambda$=4.21~MeV, which is close to  the observed
$B_\Lambda$=4.18$\pm$0.10~MeV.
Our results for (1$^-$,2$^-$) energy spacing differ from that obtained within the model of Ref. \cite{HOKY14}. At the same time,  2$^-$ resonance energies are similar.

\section{Conclusions}

The three-body  $\alpha \Lambda$n cluster model 
with the OBE simulating (NSC97f) potential for the $\Lambda \rm n$ interaction was
applied for calculation of  spin (1$^-$,2$^-$) doublet energies of the  $^6_\Lambda$He hypernucleus. We obtained the value of 0.18~MeV for the (1$^-$,2$^-$) energy spacing.
The calculated $2^-$ excitation energy is in a good agreement with the results of other calculations whereas
 ${1}^{-} $ binding energy disagrees with these calculations.
The   results obtained for the bound state energy is lower than
the experimental value. It may be an evidence for the
violation of the exact three-body cluster structure of the system.
The three-body force is required to describe experimental data
within this cluster model.
 

\end{document}